\documentclass[a4paper,11pt]{article}
\usepackage[english]{babel}
\usepackage[latin3]{inputenc}
\usepackage[T1]{fontenc}
\usepackage{textcomp}
\usepackage{epsfig}
\usepackage{latexsym}
\usepackage{graphicx}
\usepackage{amsfonts,amssymb,bm}
\usepackage{color}
\newtheorem{definition}{Definition}

\newtheorem{postulate}{Postulate}

\newcounter{llista}

\begin{document}
\title{Yet another derivation of special relativity transformations without the second postulate}
\author{Josep Llosa\\
\small Departament de Física Fonamental, Universitat de Barcelona, Spain}

\maketitle

\begin{abstract}
The most general transformation connecting inertial frames is derived from rather general and simple assumptions, without the postulate of constancy of the speed of light in vacuo.

\noindent
\end{abstract}

\section{Introduction\label{S0}}
It seems that  Einstein's second postulate, i. e. \guillemotleft in empty space light is always propagated with a definite velocity which is independent of the state of motion of the emitting body\guillemotright, has a most relevant role in the derivation of Lorentz transformation of coordinates. This is clear in his seminal paper of 1905 \cite{collected} and also in most textbooks on special relativity,

The setting of an operational method to synchronize distant clocks is indeed an important piece of the introduction in the 1905 paper. The result is what some authors call the {\em Einstein-Poincar\'e criterion of simultaneity} \cite{Galison},\cite{Gourg2010} \footnote{That protocol is related to the correction of the transmission time that had been used by telegraphic-chartographers since long ago.} 
which is based on the assumption that light propagates at the same speed in all directions.

In spite that this way of presenting things suggests that the validity of Lorentz transformations depends on this synchronization method and hence on Einstein's second postulate, it is well known that other synchronization procedures, e. g. slow transportation of previously syncronized clocks at the origin, yield the same coordinate transformations. 

Furthermore many experiments dessigned to measure relativistic effects do not depend on such a clock syncronization at all. It is so, for instance, in Frisch and Smith experiment \cite{Frisch63} intended to reveal time dilation in the mean lifetime of muons produced by cosmic rays, by measuring how many of them have desintegrated along a 2000 m trip (from the top of Mt. Washington to sea level) traveling at almost the speed of light.
Instead of measuring the speed of muons by the standard kinematical method, namely the quotient of the distance over the difference of times read by two synchronized clocks, it is determined by a dynamical effect: the muons penetration power. 
This involves no explicit synchronization of distant clocks; however, the experiment is based on the tacit assumption that muons, being practically free (the duration of the flight is too short for gravity to have an appreciable effect), move uniformly in a straight line in agreement with the first law of mechanics.

It is thus Galilei law of inertia which actually selects some synchronization protocols and rejects others. Indeed, if we tried to study the motion of a supposedly free point mass, by doing the appropriate kinematical measurements with the help of a reference space and a team of local clocks, whatever they are synchronized, we should find out that not all synchronization methods would yield the uniform motion stated by the law of inertia

The attempts to derive the special relativistic transformations of coordinates without the second postulate started more than one century ago. On the basis of the principle of relativity, homogeneity and isotropy of space, the so called principle of reciprocity ---if $\mathcal{K}$ moves at the velocity $q$ with respect to $\mathcal{K}^\prime$, then  $\mathcal{K}^\prime$ moves at $-q$ with respect to  $\mathcal{K}$--- and some kind of comparison of the standards of length between two observers in relative motion, Ignatowski\cite{Ignatowski1910} obtained that the relativistic transformations are either Galilei transformations or a family of Lorentz-like transformations that depend on a parameter $\alpha$ whose dimensions are the inverse square of a velocity. The identification of $\alpha$ with the inverse of the velocity of light, $\alpha = c^{-2}$, then follows from the length contraction factor (as observed in the deformation of the equipotential surfaces). 
A similar result, using group theory, was obtained by Frank and Rothe \cite{Frank1911}.

In his celebrated article on the theory of relativity Pauli \cite{Pauli} underestimates these results because they eventually need some additional postulate on light or on electrodynamics in order to fix the value of the unknown parameter $\alpha$.
Even though Pauli is right in that \guillemotleft 
From the group theoretical assumption it is only possible to derive the general form of the transformation formulae, but not their physical content\guillemotright,
those alternative derivations of relativity transformations make the light postulate innecessary, as far as $\alpha$ could be determined by measuring the time dilation factor, e. g. with an experiment like the aforementioned \cite{Frisch63}.

Since then many derivations of relativity transformations without the second postulate have appeared in the literature. A non-exhaustive list would include refs. \cite{Pars21} to  \cite{Stepanov2010}, to cite a few. All of them are based on some often imprecise form of the principle of relativity; e. g. in ref. \cite{Levy76}, \guillemotleft  \ldots a class of reference frames in spacetime which are physically equivalent \guillemotright, meaning that \guillemotleft  the laws of physics take the same form when referred to one of these frames\guillemotright. (Which raises the question of whether these ``laws of physics'' include electromagnetism.) 

All these derivations have to resort to some additional assumptions, that are also present in standard derivations of Lorentz trasnformations. To list a few: 
\begin{list}
{(\alph{llista})}{\usecounter{llista}}
\item the reciprocity principle (refs. \cite{Ignatowski1910}, \cite{Gorini68}, \cite{Mermin84}, \cite{Feigenbaum2008}, \cite{Stepanov2010}, \cite{Manida2008}), 
\item the homogeneity and isotropy of space (from which some wrongly infer that the transformations must be linear), \item if two particles move at the same velocity with respect to $\mathcal{K}$,  then they move at the same velocity with respect to $\mathcal{K}^\prime$ (refs. \cite{Feigenbaum2008} and  \cite{Stepanov2010}),
\item some assumptions concerning the relative veocity that  tacitly presume that both observers have previously come to an agreement concerning their respective standard rods and clocks (refs. \cite{Pars21}, \cite{Lee75} and \cite{Rind77}).
\end{list}
In our view some of these assumptions are too ``elaborated'' and do not exhibit the primary quality of ``self-evident truths'' that one would expect, similar to what one finds, for instance, in the axioms of Euclides. 

As for the reciprocity principle, not only is arbitrary but quite innecessary. Actually Levy-Leblond \cite{Levy76} obtains it from more basic assumptions for one-dimensional space.

Besides, some of the above derivations of relativity transformations are one-dimensional in space and their generalisations to three dimensions are based on the choice of two ``parallel'' Cartesian triads of axes, one in each reference frame, such that one axis in each triad, say the $X$-direction, is oriented along the direction of relative motion. However, as these triads of axes are in relative motion, the geometrical meaning of that notion of ``parallelism'' on which an important part of the proof rests is not clear at all \footnote{Indeed, does this ``parallelism'' refer to ``simultaneous'' images of the two triads of axes? As far as the second postulate and clock synchronisation is avoided, it is very unclear what does this mean.}.  

On its turn, apart from being one-dimensional in space, Levy-Leblond's derivation \cite{Levy76} involves a sort of ``counting the parameters'' of the group of inertial transformations. Although the wanted result is reached, the basis of his argument looks rather artificial. In the present work we need not to make any assumption on the number of group parameters and rely instead on our Postulate \ref{P4}, which looks simpler and more intuitive.

The fact that Poincaré-like transformations (or inhomogeneous Lorentz-like) can be derived on the basis of a set of simple general postulates, with no reference to a specific physical phenomenon, namely electromagnetism or light, seems important enough to justify the job of clarifying and simplifying these postulates as much as possible ---stating them in an intuitive way and expressing them in the everyday language--- and marking the specific role of each postulate in the derivation of inertial transformations. This has been the animating idea of the way the present work has been written.

The mathematical tools necessary to cover the steps from the hypothesis to the results are rather simple for a senior undergraduate student in physics: matrix algebra and some differential calculus. We have purposely avoided graphical reasonings with cartesian axes and statements like: ``If $\mathcal{K}_1$ sees $\mathcal{K}_2$ `like this', then, by way of reciprocity, $\mathcal{K}_2$ must see $\mathcal{K}_1$ `like that' \ldots''. These graphical methods are an important part of the physicists' art and craft, they are useful tools to illustrate interesting results in a simple and elegant way and, as such, they are frequent in textbooks. However they are often the vehicle by which tacit sssumptions slip into a proof, so blurring the relationship between a result and the postulates that support it.

The fundamental basis on which our reasoning rests is Galilei law of inertia, namely 
\guillemotleft a mass point, not acted by anything external, remains at rest or keeps in a state of uniform rectilinear motion\guillemotright, and the list of postulates is the following: 
\begin{description}
\item[Postulate 1] (A minimal principle of relativity) There is a continuous, infinite class of reference frames, namely {\em inertial frames}, in which Galilei law of inertia holds. 
\item[Postulate 2] The stationary space of an inertial frame is three-dimensional Euclidean space.
\item[Postulate 3] If an inertial frame is at rest relatively to a second inertial frame, then the transformation relating their respective systems of coordinates is an Euclidian transformation, i. e. a spacetime traslation followed by an orthogonal space transformation. 
\item[Postulate 4] If two inertial frames move at the same finite velocity with respect to a third inertial frame, then they are at rest with respect to each other.
\item[Postulate 5] It exists an invariant time ordering of events which is invariant by inertial transformations. 
\end{description}

As the contents of some of these postulates depends on definitions, like relative rest or relative velocity, which require notions derived from previous postulates, we shall further along the following sections.

Postulates 3 and 4 will eventually imply that inertial transformations are linear and depend on ten parameters, namely four for spacetime translations, three for rotations and three for the relative velocity (that we shall define more precisely further on). On its turn, Postulate 5 is necessary in order that any cause always precedes its effects for all inertial frames, so that there is a causality which is invariant by inertial transformations.

\section{The principle of relativity \label{S1}}
Any observer describes reality in terms of {\em events} and coordinates each event with the help of: (a) the {\em place} where it happens, i. e. a point in his {\em stationary space}, and (b) a {\em time} coordinate. (A means to determine the time usually consists of a set of stationary clocks, but we shall make no explicit assumption on either their rates or their synchronization.)
The stationary space of an observer together with his set of stationary clocks conform what we call his {\em reference frame}.

By the principle of relativity the laws of physics ---whatever these mean--- take the same form for a wide class of observers which we usually call {\em inertial observers}. We shall assume that Newton's first law of dynamics, i. e. the {\em Galilei law of inertia}, is included among these invariant laws. 
\begin{postulate}  \label{P1}
There is a continuous, infinite class of reference frames, namely {\em inertial frames}, in which Galilei law of inertia holds.
\end{postulate} 

According to this law, 
\guillemotleft a mass point, not acted by anything external, remains at rest or keeps in a state of uniform rectilinear motion\guillemotright. 
It is implicit in this statement that: (a) the stationary space of an inertial observer has a distinguished class of lines, namely straight lines, and (b) the notions of distance and time assignement are somehow defined, as far as uniform motion means ``equal distances in equal times''.

To make things as simple as possible, while giving content to Galilei law of inertia, we shall assume that 
\begin{postulate} \label{P2}
The stationary space of any inertial observer is Euclidean three-dimensional space.
\end{postulate} 
Space measurements are ruled by Euclidean geometry and, by a suitable choice of an origin and a triad of orthogonal axes, three Cartesian coordinates $(x^1,x^2,x^3)$ can be assigned to each point in space. 

\begin{definition} \label{D1}
An inertial system of coordinates  $\mathcal{K}$ assigns a 4-tuple of real numbers $(x^1,x^2,x^3,t)$ to each event.
\end{definition}
The first three numbers are the Cartesian coordinates of the event's place in the reference space, while the fourth corresponds to time and is assigned in a way that we don't need to especify but, as said before, the validity of Galilei law imposes a strong constraint on the way that cloks are synchronized and on their rates.

\begin{definition}   \label{D2}
An inertial transformation is a map
$$ (x^1,x^2,x^3,t) \,\stackrel{\Phi}{\longrightarrow} \, (x^{\prime 1},x^{\prime 2},x^{\prime 3}, t^\prime)  $$ 
that relates the coordinates assigned by two different inertial systems, $\mathcal{K}$ and $\mathcal{K}^\prime$, to the same event.
\end{definition}

An immediate consequence of this definition is that the class of all inertial transformations form a group. Indeed, let 
$\mathcal{K}_a$, $a=1,2,3$, be three systems of inertial coordinates and let $\Phi_{ab}$ denote the inertial transformation from $\mathcal{K}_a$ to $\mathcal{K}_b$, then the product $\Phi_{12} \circ \Phi_{23}=\Phi_{13}$ is also an inertial transformation. Besides, the inertial transformation $\Phi_{aa}$ is the identity and the inverse transformation of $\Phi_{ab}$ is $\Phi_{ba}$.

Explicit expressions for the components of an inertial transformation are
\begin{equation}  \label{e1} 
x^{\prime i} = f^i(x^j,t)\,, \qquad t^\prime = g(x^j,t)\,, \qquad \qquad i,j=1\ldots 3\,,
\end{equation} 
Now, by Postulate 1, an inertial transformation must convert any uniform rectilinear motion into a uniform rectilinear motion
$$ (x^j + t v^j,t) \, \longrightarrow \, (x^{\prime i} + t^\prime  v^{\prime i},t^\prime) \,, $$
that is:
\begin{equation}  \label{e2}
 f^i (x^j + t v^j,t) = X^i(x^j,v^k) + g(x^j + t v^j,t)\, V^i(x^j,v^k) \,, \qquad \forall x^j,\, t \,, \qquad
i,j=1 \ldots 3 \,,
\end{equation}
where $v^k$ is the velocity with respect to the inertial system of coordinates $\mathcal{K}$ and $V^i(x^j,v^k)$ is the velocity with respect to $\mathcal{K}^\prime$.

It is well known ---see for instance refs.  \cite{Fock} and \cite{Aharoni85}--- that the most general transformation fulfilling condition (\ref{e2}) is
\begin{equation}  \label{e3}
 f^i (x^j,t) = \frac{A^i_{\;j} \, x^j + c^i \,t + a^i}{h_i x^i + k t + l} \,, \qquad \qquad g(x^j,t) = \frac{ b_j\, x^j + \gamma \, t + e }{h_i x^i + k t + l} \,, \qquad
\end{equation}
where summation over repeated indices is understood.

This is a transformation of four dimensional real projective space and is called a {\em homography} \cite{projectivity}. 
There is a correspondence between these transformations and the class of $5D$ special matrices $\mathbb{A} \in SL(5,\mathbb{R})$; indeed, equations (\ref{e3}) can be written as:
\begin{equation}  \label{e4}
 X^\prime = \frac1{f(X)} \,\mathbb{A}\,X
\end{equation}
where
\begin{equation}  \label{e5}
 X^\prime :=  \left( \begin{array}{c}
            \vec{x}^\prime \\ \hline t^\prime \\ \hline 1 \end{array}  \right) \,, \qquad
 X :=  \left( \begin{array}{c}
            \vec{x} \\ \hline t \\ \hline 1 \end{array}  \right) \,, \qquad
 \mathbb{A} := \left( \begin{array}{c|c|c}
                     \; \mathbf{A} \; & \vec{c}  & \vec{c}\\
             \hline \; \vec{b}^T & \gamma & e \\ 
             \hline \; \vec{h}^T & k & l
                     \end{array}  \right) \,, \qquad {\rm with }
\end{equation}
$$\det\mathbb{A}=1 \qquad \qquad {\rm and }  \qquad \qquad f(X) := \vec{h}\cdot \vec{x}+ k t + l$$ 
(a ``dot'' meaning the ordinary scalar product of two 3-vectors).

Furthermore, this correspondence is one-to-one and it can be easily checked that the product of two homographies, with associated matrices $\tilde{\mathbb{A}}$ and $\mathbb{A}\,$, is also a homography whose associated matrix is the product $\tilde{\mathbb{A}}\, \mathbb{A}$.

We have seen so far that, as a consequence of the minimal principle of relativity alone, the class of inertial transformations is a subgroup of the group of four-dimensional projectivities.

\section{Relative velocity  \label{S2}}
The ``history´´ of a given place $P$ in the stationary space of an inertial observer $ \mathcal{K}$ is a 1-dimensional continuum of events that, in the coordinates of $ \mathcal{K}$, is
$\,  (\vec x,\,t) =  (x^1,\,x^2,\,x^3,\,t) \,, \quad  t \in \mathbb{R} \, $.

In a different system of inertial coordinates, the world-line of $P$ is obtained by the coordinate transformation (\ref{e3}) and we have that, according to $\mathcal{K}^\prime$, $P$ moves at the velocity:
\begin{equation}   \label{e7}
 \vec{u}(\vec{x}) = \frac{\vec{x}^\prime(\vec{x},t) - \vec{x}^\prime(\vec{x},0)}{t^\prime(\vec{x},t) - t^\prime(\vec{x},t)}  = \frac{(l+\vec{h}\cdot\vec{x})\,\vec{c} - k \,\left(\mathbf{A}\vec{x} + \vec{a} \right)}{(l+\vec{h}\cdot\vec{x})\,\gamma - k \,(\vec{b}\cdot\vec{x} + e )}
\end{equation}
(here $\mathbf{A}\vec{x}$ means the product of a square matrix by a column vector to yield a column vector). This velocity depends on $\vec x$ and, as a rule, different points in the $\mathcal{K}$-space could have different velocities as  seen by $\mathcal{K}^\prime$.

\begin{definition}   \label{D3}
The {\em relative velocity} of $\mathcal{K}$ with respect to $\mathcal{K}^\prime$ is the velocity of the origin of $\mathcal{K}$, i. e. $\vec u(0)$, 
\begin{equation}  \label{e8}
\vec{u}:=  \frac{l\,\vec{c} - k \, \vec{a}}{l \gamma - k e} \,.
\end{equation}
If $l \gamma - k e =0$, we say that the relative velocity is infinite.
\end{definition}
In case that the relative velocity is infinite, $\mathcal{K}^\prime$ sees the origin of $\mathcal{K}$ as being simultaneously in every point of a straight line in the $\mathcal{K}^\prime$-space.

When $\vec u = 0$ we say that $\mathcal{K}^\prime$ is {\em at rest} with respect to $\mathcal{K}$. 

The next postulate expresses the intuitive fact that to observers at relative rest can use the same equipment of stationary clocks and the same standard length in the conventional way that is understood in Euclidean geometry. 
\begin{postulate}  \label{P3}
If the inertial system $\mathcal{K}$ is at rest with respect to $\mathcal{K}^\prime$, then the transformation connecting them is an Euclidian transformation, i. e. a spacetime translation followed by a space orthogonal transformation.
\end{postulate}

By this postulate, the $5D$ matrix representing the inertial transformation from $\mathcal{K}$-coordinates to $\mathcal{K}^\prime$-coordinates is 
\begin{equation}  \label{e9}
\mathbb{S} = \left( \begin{array}{c|c|c}
                     \; \mathbf{S} \; & 0  & \vec m \\
             \hline \; \vec{0}^T & 1 & n \\
             \hline \; \vec{0}^T & 0 & 1    \end{array}  \right) 
\end{equation}
where $\mathbf{S} \in O(3)$ is an orthogonal matrix, $\vec m$ is a 3-vector and $n$ a real number. We shall call the $5D$ matrix $\mathbb{S}$,  a {\em rest matrix} and, as the corresponding factor $f(X)$ in equation (\ref{e4}) is 1, we have that $X^\prime =\mathbb{S} \,X$.

The fact that a Euclidean transformation is the result of a spacetime translation followed by a space orthogonal transformation, can be expressed in terms of matrices as:
\begin{equation}  \label{e9a}
 \mathbb{S} = \mathbb{S}_0 \,\mathbb{T}
\end{equation}
where 
$$ \mathbb{S}_0 = \left( \begin{array}{c|c|c}
                     \; \mathbf{S} \; & 0  & 0 \\
             \hline \; \vec{0}^T & 1 & 0 \\
             \hline \; \vec{0}^T & 0 & 1    \end{array}  \right) \,, \qquad \qquad 
   \mathbb{T} = \left( \begin{array}{c|c|c}
             \; \mathbf{1}_3 \; & 0  & \vec m \\
             \hline \; \vec{0}^T & 1 & n \\
             \hline \; \vec{0}^T & 0 & 1    \end{array}  \right)  \,, $$
that is, a rest matrix is the product of a {\em homogeneous rest matrix} $\mathbb{S}_0$ times a {\em translation matrix} $\mathbb{T}$.

It is also easy to see that the inverse of a rest matrix is also a rest matrix, indeed,
$$ \mathbb{S}^{-1} = \left( \begin{array}{c|c|c}
                     \; \mathbf{S}^T \; & 0  & - \mathbf{S}^T \vec m \\
             \hline \; \vec{0}^T & 1 & - n \\
             \hline \; \vec{0}^T & 0 & 1    \end{array}  \right)   $$
Therefore, if $\mathcal{K}$ is at rest with respect to $\mathcal{K}^\prime$, then the converse is also true, and the relation of being at rest is mutual.

The next postulate also expresses an intuitive notion:
\begin{postulate}  \label{P4}
Two inertial frames, $\mathcal{K}^\prime$ and $\mathcal{K}^{\prime\prime}$, move at the same finite velocity with respect to a third inertial frame $\mathcal{K}$ if, and only if, $\mathcal{K}^\prime$ and $\mathcal{K}^{\prime\prime}$
are at rest with respect to each other.
\end{postulate}

In terms of the associated $5D$ matrices $\mathbb{A}$ and $\tilde{\mathbb{A}}$, the coordinate transformations from $\mathcal{K}^\prime$ to $\mathcal{K}$ and from $\mathcal{K}^{\prime\prime}$ to $\mathcal{K}$ are, respectively,
\begin{equation}  \label{e10}
X = \frac1{f(X^\prime)}\, \mathbb{A} X^\prime \qquad {\rm and} \qquad 
X = \frac1{\tilde{f}(X^{\prime\prime})}\, \tilde{\mathbb{A}} X^{\prime\prime}
\end{equation}
with
\begin{equation}  \label{e11} 
\mathbb{A} := \left( \begin{array}{c|c|c}
                     \; \mathbf{A} \; & \vec{c}  & \vec{a}\\
             \hline \; \vec{b}^T & \gamma & e \\ 
             \hline \; \vec{h}^T & k & l
                     \end{array}  \right) \,, \qquad \qquad 
 \tilde{\mathbb{A}} := \left( \begin{array}{c|c|c}
                     \; \tilde{\mathbf{A}} \; & \tilde{\vec{c}}  & \tilde{\vec{a}}\\
             \hline \; \tilde{\vec{b}}^T & \tilde{\gamma} & \tilde{e} \\ 
             \hline \; \tilde{\vec{h}}^T & \tilde{k} & \tilde{l}
                     \end{array}  \right)  
\end{equation}
and 
$$  f(X^\prime) := \vec{h}\cdot \vec{x}^\prime+ k t^\prime + l \,, \qquad \qquad \tilde{f}(X^{\prime\prime}) := \tilde{\vec{h}}\cdot \vec{x}^{\prime\prime}+ \tilde{k} t^{\prime\prime} + \tilde{l}  $$

According to the definition \ref{D3}, the relative velocities of $\mathcal{K}^\prime$ and $\mathcal{K}^{\prime\prime}$ with respect to $\mathcal{K}$ are, respectively, 
\begin{equation}  \label{e13}
\vec{u} = \frac{l\,\vec{c} - k \, \vec{a}}{l \gamma - k e}  \qquad {\rm and}  \qquad 
\tilde{\vec{u}} = \frac{\tilde{l}\,\tilde{\vec{c}} - \tilde{k} \, \tilde{\vec{a}} }{\tilde{l} \tilde{\gamma} - \tilde{k} \tilde{e}} \,,
\end{equation}
and they are finite if, and only if, $\left(l \gamma - k e \right) \left(\tilde{l} \tilde{\gamma} - \tilde{k} \tilde{e}\right) \neq 0\,$.

By Postulate \ref{P3}, the fact that  $\mathcal{K}^\prime$ and $\mathcal{K}^{\prime\prime}$ are mutually at rest
means that there exists a rest matrix $\mathbb{S}$ such that $X^\prime = \mathbb{S} X^{\prime\prime}$. Combining this with equation (\ref{e10}), we obtain that 
$$ \frac1f\,\mathbb{A} \mathbb{S} X^{\prime\prime} = \frac1{\tilde{f}}\,\tilde{\mathbb{A}} X^{\prime\prime} \,, \qquad \forall X^{\prime\prime} = \left( \begin{array}{c}
            \vec{x}^{\prime\prime} \\ \hline t^{\prime\prime} \\ \hline 1 \end{array}  \right) \,, $$
which, including that $\det\mathbb{A} = \det\tilde{\mathbb{A}}=\det\mathbb{S}=1$, implies that
\begin{equation}  \label{e14}
  \mathbb{A} \mathbb{S}  = \tilde{\mathbb{A}}
\end{equation}
or, expliciting the several blocks,
\begin{eqnarray}  \label{e15}
\tilde{\mathbf{A}} = \mathbf{A} \mathbf{S} \,, \qquad &  \qquad \tilde{\vec c} = \vec c\,, \qquad & \qquad \tilde{\vec a} = \vec a + \mathbf{A} \vec m + n \vec c \\   \label{e16}
\tilde{\vec b} = \mathbf{S}^T \vec b \,, \qquad &  \qquad \tilde{\gamma} = \gamma\,, \qquad & \qquad \tilde{e} = e + \vec b\cdot \vec m + n \gamma \\   \label{e17}
\tilde{\vec h} = \mathbf{S}^T \vec h \,, \qquad &  \qquad \tilde{k} = k \,, \qquad & \qquad \tilde{l} = l + \vec h\cdot \vec m + n k
\end{eqnarray}  

Therefore, Postulate \ref{P4} amounts to 
\begin{equation}  \label{e18}
\tilde{\vec{u}} = \vec{u} \quad {\rm finite}) \qquad \Leftrightarrow \qquad \mbox{it exists a rest matrix}\quad \mathbb{S} \;\;\mbox{such that} \quad \tilde{\mathbb{A}} = \mathbb{A} \mathbb{S} 
\end{equation}

The ``if'' ($\Leftarrow$) implication easily leads to:
\begin{equation}  \label{e19}
  k\,\left( [\gamma l - k e]\,\mathbf{A} + \vec{a}\,[k \vec{b}^T - \gamma \vec{h}^T] - \vec{c}\,[l \vec{b}^T - e \vec{h}^T]  \right)\, \vec{m} = 0 \,, \qquad \forall \vec m \in \mathbb{R}^3\,,
\end{equation}
where the expressions (\ref{e13}) to (\ref{e17}) have been included (some terms in the left hand side, as $\vec a\,\vec b^T$, must be read as the square matrix product of column times row matrix). 

If $k\neq 0$, the $3D$ square matrix between the brackets in the left hand side must vanish, which eventually implies that $\det \mathbb{A}=0$. Therefore the solution to equation (\ref{e19}) is $k=0$ and then the condition of finite velocity reduces to $l \tilde{l} \gamma \neq 0$.

Let us now substitute this $k=\tilde k = 0$ in equations (\ref{e11}). The ``only if'' ($\Rightarrow$) implication 
in the statement (\ref{e18}) then means that, whenever $ \displaystyle{\frac{\tilde{\vec{c}}}{\tilde\gamma} = \frac{\vec{c}}{ \gamma} = \vec u }\,$, it exists a rest matrix $\,\mathbb{S} \,$ such that  $\,\tilde{\mathbb{A}} = \mathbb{A} \mathbb{S}\,$. 

Now, by the polar decomposition theorem \cite{Miller}, given the $3D$ matrix $\mathbf{A}$, it exists an orthogonal matrix $\mathbf{W} \in O(3)$ such that
\begin{equation}  \label{e21} 
\mathbf{A} \mathbf{W} = \mathbf{H} \qquad \mbox{symmetric and non-negative.}
\end{equation}
(Besides, if $\det\mathbf{A} \neq 0$, this orthogonal matrix is unique.)

Let us now take the homogeneous rest matrix 
$$ \mathbb{W}_0 := \left( \begin{array}{c|c|c}
                     \; \mathbf{W} \; & 0  & 0 \\
             \hline \; 0 & 1 & 0 \\
             \hline \; 0 & 0 & 1
                     \end{array}  \right)  $$
and define the {\em boost} matrix associated to $\mathbb{A}$ as
\begin{equation}  \label{e22} 
\mathbb{B}:=\mathbb{A} \mathbb{W}_0 =  \left( \begin{array}{c|c|c}
                                        \; \mathbf{H} \; & \vec{c}  & \vec{a}\\
                                        \hline \; \vec{w}^T & \gamma & e \\ 
                                        \hline \; \vec{p}^T & 0 & l
                                        \end{array}  \right) \,, 
\end{equation}
where $\vec p := \mathbf{W}^T \vec h$ and $\vec w := \mathbf{W}^T \vec b$. (According to this definition, a boost matrix is characterized by the facts that: (a) its $3D$ spatial block is a symmetric non-negative matrix and (b) $\mathbb{B}^4_{\;5}=0$.)

Now, by the first of equations (\ref{e15}) and including equation (\ref{e21}), we easily obtain 
Therefore the homogeneous rest matrix $\tilde{\mathbb{W}}_0 = \mathbb{S}_0^T \mathbb{W}_0$ can be used to obtain the boost associated to $\tilde{\mathbb{A}}$. Indeed, writing $\mathbb{S}= \mathbb{S}_0 \mathbb{T}\,$, as in the decomposition (\ref{e9a}), we have
$$ \tilde{\mathbb{B}} = \tilde{\mathbb{A}} \tilde{\mathbb{W}}_0 = 
\mathbb{B} \mathbb{W}^T_0 \mathbb{S}_0 \mathbb{T} \mathbb{S}_0^T \mathbb{W}_0 $$
It is straightforward that $\tilde{\mathbb{T}} := \mathbf{W}^T_0 \mathbb{S}_0 \mathbb{T} \mathbb{S}_0^T \mathbf{W}_0 $ is also a translation matrix and therefore
\begin{equation}  \label{e24} 
\tilde{\mathbb{B}}:=\mathbb{B} \tilde{\mathbb{T}} =  \left( \begin{array}{c|c|c}
                                        \; \mathbf{H} \; & \vec{c}  & \tilde{\vec{a}}\\
                                        \hline \; \vec{w}^T & \gamma & \tilde e \\ 
                                        \hline \; \vec{p}^T & 0 & \tilde l
                                        \end{array}  \right) \,, 
\end{equation}
with $\vec c = \gamma \vec u$ and 
$$ \tilde{\vec{a}} = \vec a + \mathbf{H} \vec m + n \gamma \vec u \,, \qquad
\tilde e = e + \vec w^T \mathbf{S} \vec m + n \gamma \,, \qquad
\tilde l = l + \vec p^T \mathbf{S} \vec m    $$
Notice that the first four columns of the boosts  $\mathbb{B}$ and $\tilde{\mathbb{B}}$ are equal and therefore $\mathbf{H}$, $\vec w$, $\vec p$ and $\gamma$ only depend on the relative velocity $\vec u$. 

To sumarize, an inertial transformation $\mathbb{A}$ relating the coordinates of two inertial frames, $\mathcal{K}^\prime$ and $\mathcal{K}^{\prime\prime}$, moving at the same velocity $\vec u$ with respect to $\mathcal{K}$, can be decomposed as:
$$ \mathbb{A} = \mathbb{B} \,\mathbb{W}^T_0 $$
where $\mathbb{B}$ is a boost for the relative velocity $u^i = \mathbb{B}^i_{\;4}/\mathbb{B}^4_{\;4}\,$, $i=1\ldots 3$, and $\mathbb{W}_0$ is a homogeneous rest transformation.

\section{The group property  \label{S4} }
We shall now use the group property of inertial transformations to obtain the different elements in the columns 1 to 4 of the boost matrix $\mathbb{B}$ as functions of $\vec{u}$.
 
For any given homogeneous rest matrix $\mathbb{S}_0$  and any boost $\mathbb{B}$, the matrix $\mathbb{S}_0\,\mathbb{B}\,\mathbb{S}^T_0$ also belongs to the group of inertial transformations and, as such, it admits a polar decomposition. Therefore they exist a boost $\mathbb{B}^\prime$ with relative velocity $\vec{u}^\prime$ and a rest matrix $\mathbb{U}$ such that
\begin{equation}  \label{e25}
  \mathbb{S}_0\,\mathbb{B}\,\mathbb{S}_0^{-1} = \mathbb{B}^\prime \,\mathbb{U}
\end{equation}
which, split in the usual spacetime blocks and after some algebra, leads to
\begin{equation}  \label{e26}
 \gamma^\prime = \gamma \,, \qquad \vec{u}^\prime = \mathbf{S}\,\vec{u} \,, \qquad \vec{w}^\prime = \mathbf{S}\,\vec{w}  \,, \qquad \vec{p}^\prime = \mathbf{S}\,\vec{p} \,, \qquad 
\mathbf{H}^\prime  = \mathbf{S}\,\mathbf{H}\,\mathbf{S}^T
\end{equation}
Now, as $\mathbf{S}\,\mathbf{H}\,\mathbf{S}^T$ is already a non-negative symmetric matrix, it follows that  $\mathbb{U} = \mathbb{I}_5$.

Thus, by an orthogonal transformation $\mathbf{S}$ of the variable $\vec{u}$, $\mathbf{H}(\vec{u})$ transform as a symmetric 2-tensor, $\vec{w}(\vec{u})$   and   $\vec{p}(\vec{u})$ as vectors  and $\gamma(\vec{u})$ as a scalar and, therefore
\begin{equation}  \label{e27}
\mathbb{B} =  \left( \begin{array}{c|c|c}
                   \; f(u) \,\mathbf{1}_3 + g(u)\, \hat{u} \,\hat{u}^T  \; & \gamma(u) \vec{u}  & \vec{a}\\
            \hline \; \mu(u) \hat{u}^T & \gamma(u) & e \\ 
            \hline \; \lambda(u) \hat{u}^T & 0 & l
                   \end{array}  \right) \,, 
\end{equation}
where $\hat{u}$ is the unitary vector and $\vec{u} = u \,\hat{u}$. 

Notice that equation (\ref{e25}) with $\mathbf{S}=-\mathbf{1}_3$ implies that 
$$\mathbf{H}(-\vec{u})= \mathbf{H}(\vec{u}) \,, \qquad 
\vec{w}\left(-\vec{u}\right) = - \vec{w}(\vec{u}) \,, \qquad 
\vec{p}\left(-\vec{u}\right) = - \vec{p}(\vec{u}) $$ 
and hence the  functions $f(u)$, $g(u)$ and $\gamma(u)$ are even whereas $\mu(u)$ and $\lambda(u)$ are odd.
Including the fact that for a rest matrix $\mathbf{H}(\vec{0}) = \mathbf{1}_3$ and $\gamma(0)=1$, the latter implies that
\begin{equation}   \label{e28}
 f(0) = \gamma (0) = 1 \,,  \qquad g(0) = \mu(0) =  \lambda(0) =0 \,,\qquad \dot f(0) = \dot g(0) = \dot\gamma(0) = 0
\end{equation}

Moreover, the condition $\det\mathbb{B}=1$ sets the constraint 
\begin{equation}   \label{e28a}
 1 = f^2 \gamma \,\left[ l(f+g - \mu \,u) + \lambda (e u - \gamma \vec a\cdot \hat u)\right] 
\end{equation}
whence it follows that $f \neq 0\,$.

Further restrictions on the one-variable functions $f$, $g$, $\gamma $, $\mu$ and $\lambda$ follow from the group property. Indeed, consider two boost matrices $\mathbb{B}_1$ and $\mathbb{B}_2$ corresponding to parallel velocities $\vec{u} = u\, \hat u$ and $\vec{v} = v\, \hat u$. As their product is an inertial matrix, it must admit a polar decomposition 
\begin{equation}  \label{e29}
\mathbb{B}_1 \,\mathbb{B}_2 = \mathbb{B}_3\,\mathbb{W}_0 
\end{equation}
$\mathbb{B}_3$ being a boost for a velocity $\vec s(\vec u,\vec v)$ and $\,\mathbb{W}_0$ a homogeneous rest matrix.

In particular, by examining the matrix element in the fifth row and fourth column of this product, we obtain that
$$  u \lambda(v)\gamma(u) = 0  \,, \qquad \forall \,u,\,v \,,   $$
that implies  
\begin{equation}   \label{e30}
\lambda(v) = 0  \,,\qquad \forall \,v 
\end{equation}
which, substituted in equations (\ref{e27}) and (\ref{e28a}) leads to  
\begin{equation}   \label{e30a}
\mathbb{B}^5_{\;a} = l\,\delta^5_a\,,\qquad a=1\ldots 5  \,, \qquad {\rm and} \qquad 
 1 = f^2 \gamma \, l\,\left(f+g - \mu \,u \right) 
\end{equation}
and hence the inertial transformation (\ref{e3}) is linear.

Using that $\lambda(u)=0$ and examining the first four columns in the product (\ref{e29}), we also obtain that 
$\mathbf{W}=\mathbf{1}_3\,$, $\,\vec s=s(u,v)\,\hat u$ and that
\begin{eqnarray}  \label{e31a}
 & & \gamma (v)\,\left\{\left[f(u)+g(u)\right]\,v+ u\,\gamma(u)\right\} \, \hat{u} = \gamma (s)\, \vec{s}  \\ \label{e31b}
 & & \gamma (v)\,\left[\mu(u)\,v + \gamma(u)\right] = \gamma (s) \\   \label{e31c}
 & & f(u)\,f(v) = f(s) \\   \label{e31d}
 & & \left[f(u) + g(u) \right]\,\left[f(v) + g(v) \right]+ u\,\gamma(u)\,\mu(v) = f(s)+g(s) \\  \label{e28e}
 & & \left[\mu(u) \left[f(v) + g(v) \right]  + \gamma(u)\,\mu(v)\right]\,\hat{u} = \mu(s)\,\hat{s}
\end{eqnarray}

The quotient between equations (\ref{e31a}) and (\ref{e31b}) yields the law of addition for parallel velocities:
\begin{equation}  \label{e32}
s(u,v) = \frac{\left[f(u)+g(u)\right]\,v+ u\,\gamma(u)}{\mu(u)\,v + \gamma(u)}
\end{equation}
which has the obvious limit $s(u,0)=u$. 

Then, taking the partial derivatives with respect to $v$ of equations (\ref{e31b}) to (\ref{e32}) at $v=0$ and including the rest values (\ref{e28}), we obtain
\begin{equation}  \label{e33a}
 \dot f(u) \,\frac{\partial s(u,0)}{\partial v} = 0 \,, \qquad 
\dot \gamma(u)\,\frac{\partial s(u,0)}{\partial v} = \mu(u)   \,,\qquad   \dot \mu(u)\,\frac{\partial s(u,0)}{\partial v} =\gamma(u)\,K 
\end{equation}
and 
\begin{equation}  \label{e33b}
 \frac{\partial s(u,0)}{\partial v} = \frac{f(u) + g(u) - u\,\mu(u)}{\gamma(u)}   \,,\qquad 
 \left[\dot f(u)+ \dot g(u)\right]\,\frac{\partial s(u,0)}{\partial v} =u\,\gamma(u)\,K  
\end{equation}
where we have written $K:= \dot\mu(0)\,$.

Now $\displaystyle{\frac{\partial s(u,0)}{\partial v}}$ cannot vanish because otherwise equations (\ref{e33a}) and (\ref{e33b}) would imply that $\mu(u)=f(u) + g(u) =0$, in contradiction with equation (\ref{e30a}). Therefore it follows from equations (\ref{e33a}) and (\ref{e28}) that $\dot f(u)=0$ and hence $f(u)=1$.

Combining again equations (\ref{e33a}) and (\ref{e33b}) and introducing $\,\kappa(u):= 1 + g(u) - u\,\mu(u)\,$, we have that
\begin{equation}  \label{e32a}
\mu(u)\, \dot\mu(u) = \gamma(u)\, \dot\gamma(u)\, K \,, \qquad 
\kappa(u)\, \dot \gamma(u) = \mu(u)\, \gamma(u) \,, \qquad 
 \dot\kappa(u) = - \mu(u) \,,
\end{equation} 
whose solution for the initial data (\ref{e28}) is
\begin{equation}  \label{e33}
\kappa(u)= \frac1{\gamma(u)} \,, \qquad \qquad   \mu^2(u) = K\,\left(\gamma ^2(u) -1\right)
\gamma(u) = \frac1{\sqrt{1-K u^2}} 
\end{equation} 
The boost coefficients thus depend on the parameter $K:=\dot\mu(0)$ which has the dimensions of the inverse square of a velocity. 

Finally, using that $f=\kappa\gamma=1$, the relation (\ref{e30a}) leads to $l=1$ which implies that the general boost matrix is
\begin{equation}  \label{e34}
\mathbb{B} = \left( \begin{array}{c|c|c}
                       \mathbf{1}_3 + (\gamma -1)\,\hat{u}\hat{u}^T & \gamma\,\vec{u} & \vec a\\  \hline
                        K\,\gamma\,\vec{u}^T & \gamma & e \\  \hline
                        \vec 0 & 0 & 1
                        \end{array} \right)
\end{equation}
where equations (\ref{e27}), (\ref{e30}) and (\ref{e33}) have been included.
The meaning of this transformation is determined by the parameter $K$: 
\begin{list}
{{\bf (\alph{llista})}}{\usecounter{llista}}
\item If $K>0$, then $\mathbb{B}$ is a Lorentz-like inhomogeneous boost and the group of inertial transformations is a Poincaré-like group. In this case, the critical velocity $c=K^{-1/2}$ acts as the limit speed, because $\gamma(c)=\infty$ and $\gamma(u)$ is an imaginary number for $u> c\,$. Besides, $s(u,c)=c$ and therefore $c$ is the only velocity which is invariant by the law of addition of velocities (\ref{e32}).
\item If $K=0$, then $\mathbb{B}$ is a Galilei boost and the group of inertial transformations is the inhomogeneous Galilei group.
\item If $K<0$, it can be easily checked that $\mathbb{B}$ is a Euclidean transformation in four dimensions for the positive metric ${\rm diag}\,(1,1,1,|K|)$ and the structure of the group of inertial transformations is similar to the 4-dimensional Euclidean group. There is no limit on the magnitude of the relative velocity in this case.
\end{list}

The law of addition of parallel velocities (\ref{e32}) then reads
$$ s(u,v) = \frac{u+v}{1+Kuv} $$
which, in the case $K<0$ allows the composition of the two finite velocities $u$ and $v = -\frac1{Ku} $ to yield an infinite velocity. 
It is not so in the case $K=0$, that $s(u,v)= u+v$, nor in the case $K>0$ in which, as $|u|$ and $|v|$ cannot exceed $K^{-1/2}$, the denominator in $s(u,v)$ never vanishes.  

\subsection{The preservation of causality  }
The case $K<0$ can be ruled out on the basis of the preservation of causality.
\begin{definition}
An event $E$ is {\em previous} to another event $F$ if for all inertial frames $t_E < t_F\,$. 
\end{definition}
This introduces a partial time ordering between pairs of events.

\begin{postulate}  \label{P5}
There is at least a couple of events, $E$ and $F$, such that $E$ is previous to $F$.
\end{postulate}

It can be easily seen \cite{Levy76} that this assumption is violated in the case $K<0$. Indeed, if $E$ is previous to $F$, we have that:
$$ t_F^\prime - t_E^\prime = \gamma \,\left[ \,(t_F-t_E) + K \vec{u}\cdot \left(\vec{x}_F - \vec{x}_E \right) \,\right] \, $$
where the boost matrix (\ref{e34}) has been used. The sign of $t_F^\prime - t_E^\prime$ differs from the sign of $t_F-t_E$ for all relative velocities 
$$\vec{u} = - \lambda \,\left(\vec{x}_F - \vec{x}_E \right) \,, \qquad {\rm with}  \qquad \lambda > (t_F-t_E) K^{-1}\, \left|\vec{x}_F - \vec{x}_E \right|^{-2} \,, $$ 
which contradicts the assumption that $E$ is previous to $F$.

This argument only applies if $K<0$ because in this case there is no limit in the magnitude of relative velocity. In the case $K \geq 0$, on the contrary, the above argument does not apply  because, for some space separations, the magnitude of  $\vec{u}$ to produce the time sign reversal would 
exceed the limit velocity $c=K^{-1/2}$. 

\section{Inertial transformations at an infinite relative velocity \label{S5}}
Insofar we have confined to transformations with a finite relative velocity. We consider now an inertial transformation $\mathbb{A}$ such that $l \gamma - k e = 0$ that, according to Definition \ref{D3}, connects an inertial frame $\mathcal{K}$ moving at an infinite velocity with respect to $\mathcal{K}^\prime$. Of course, for this inertial matrix it must be $l\,\vec c - k\,\vec a \neq 0$, because otherwise $\det\mathbb{A} = 0\,$.

Let us take a homogeneous boost matrix $\mathbb{B}_0$ with a finite velocity $\vec v$ and consider the inertial matrix $\tilde{\mathbb{A}} = \mathbb{B}_0\mathbb{A}\,$. In order to assess whether the corresponding relative velocity is finite or infinite, we must evaluate $\tilde{l} \tilde{\gamma} - \tilde{k} \tilde{e}$, that is
$$ \tilde{l} \tilde{\gamma} - \tilde{k} \tilde{e} = K \gamma(v) \vec v\cdot \left(l\,\vec c - k\,\vec a  \right) \,.$$ 
As  $l\,\vec c - k\,\vec a \neq 0$, it exists $\vec u$ such that $\tilde{l} \tilde{\gamma} - \tilde{k} \tilde{e} \neq 0$ 
and the inertial matrix $\tilde{\mathbb{A}}$ has a finite relative velocity. Therefore, as seen in section \ref{S2}, it admits a polar decomposition 
$\tilde{\mathbb{A}} = \tilde{\mathbb{B}} \mathbb{W}_0 \,$, where $\tilde{\mathbb{B}}$ is a boost and $\mathbb{W}_0$ is a homogeneous rest matrix, whence it follows that 
$$  \mathbb{B}^{-1} \tilde{\mathbb{B}} = \mathbb{A} \mathbb{W}^T_0 $$
Now, as the matrix product by $\mathbb{W}^T_0$ does not modify the relative velocity, the inertial matrix $\mathbb{B}^{-1} \tilde{\mathbb{B}}$ must have an infinite relative velocity. But, as seen at the end of section \ref{S4}, the composition of two finite velocities is also a finite velocity, except in the case $K<O$, which has been ruled out on the basis of Postulate \ref{P5}.

\section{Conclusions and outlook}
On the basis of a simple set of rather elementary postulates ---namely the relativity principle, Euclidean space, Galilei law of inertia, on relative rest and relative motion and a causality principle--- we have delimitated the general form of inertial transformations of coordinates. Each of them is the result of: (a) a space rotation (and maybe parity), a boost (Lorentz-like or Galilei, depending on certain parameter that has the dimensions of the inverse square of a speed) and (c) a spacetime translation. 

We have thus proved that the class of inertial transformations is either a Poincaré-like group ($K>0$) or the inhomogeneous Galilei group ($K=0$). 
The proof has been worked for three space dimensions, however, as this number has played no relevant role in the proof,
the result could be extensible to any number of space dimensions as well.

Finally a word is due to comment the set of postulates we have chosen. As such, they look ``self-evident truths'' but some of them are less ``self-evident'' than others. One may think of relaxing Postulate \ref{P2} . This would require a generalization of the law of inertia, defining the class of free motions in terms of the geodesics of some spacetime connection; the least extension could consist in assuming that the reference space of any inertial frame is a constant curvature three dimensional space. 

Notice also that Postulate \ref{P3} means that, if two observers are at rest with respect to each other, they can come to an agreement concerning their standards of length and also share the same equipment of stationary clocks. Doing so they give up the freedom of differently synchronizing distant clocks. Hence the considerations raised by Mansouri and Sexl  \cite{Mansouri77} are excluded in our approach.

\section*{Acknowledgement}
The author is indebted to J Masoliver and A Molina for useful comments on the present manuscript.

\end{document}